# An Approach to Estimating Quadratic Logistic Model Parameters of Fractal Dimension Curves

Yanguang Chen

(Department of Geography, College of Urban and Environmental Sciences, Peking University, 100871, Beijing, China. Email: chenyg@pku.edu.cn)

**Abstract:** The fractal dimension curves of urban form and growth fall into two categories: One can be described by common logistic function, and the other can be described with quadratic logistic function. The approach to estimating the parameter of the ordinary logistic model has been developed. However, how to estimate the parameter of quadratic logistic model is still a problem. This paper is devoted to finding a nonlinear regressive approach for estimating parameter values of quadratic logistic model of fractal dimension curves. The process can be summarized as below. First, differentiating quadratic logistic function in theory with respect to time yields a growth rate equation of fractal dimension. Second, discretizing the growth rate equation yields a nonlinear regressive model of fractal dimension curve. Third, applying the least squares method to the nonlinear regressive equation yields the capacity parameter value of the quadratic logistic model. Fourth, substituting the capacity parameter value into the quadratic logistic model and changing it into a quasilinear form, we can estimate the other parameter values by ordinary linear regression analysis. In this way, a practical quadratic logistic model of fractal dimension curves can be gained. The approach is applied to multifractal dimension curves of Beijing city to show its effectiveness. The method can be extended to estimate the parameter values of quadratic logistic models in many fields besides urban science.

# 1 Introduction

The form and growth of cities in different countries have both similarities and differences. The commonality lies in the fact that urban morphology has fractal characteristics, and fractal dimension time series exhibit squeezing effects (Batty and Longley, 1994; Benguigui *et al*, 2000; Chen and Huang, 2019; Shen, 2002). The time series of fractal parameters can be described with sigmoid functions. The difference is that fractal growth of some cities exhibit a conventional logistic curve (Chen, 2012; Man and Chen, 2020), while the fractal dimension curves of other cities exhibit a quadratic logistic curve (Chen, 2018; Chen and Huang, 2019). Fractal dimension curves of urban form can be used to predict urban growth, identify urban development stages, find influence factors of urban evolution, and explore urban dynamics. The prerequisite is to establish a proper logistic model for urban growth based on the fractal dimension of urban form and effectively estimate its parameter values. For the common logistic model of fractal dimension curve of cities, one of desirable methods of parameter estimation is to combine ordinary least squares calculation with nonlinear autoregressive analysis. However, so far, there is no simple and feasible way to estimate parameters of the quadratic logistic model of urban growth.

The key step in model construction, whether for ordinary logistic functions or quadratic logistic functions, is to determine the capacity parameter values. For the logistic models of fractal dimension curves of urban form, the capacity parameter is the largest value of fractal dimension in future. In practice, the capacity parameters of the model can be estimated using the brute force search method. However, for researchers who are not familiar with mathematical calculations, the brute force search method is not much applicable. Moreover, brute force search methods sometimes overestimate capacity parameter values. In this paper, a new approach is developed and proposed to estimate the parameter values of quadratic logistic model of fractal dimension curve of urban forma and growth. The method can be extended to other fields involving quadratic logistic modeling. The rest parts are arranged as follows. In Section 2, three sets of mathematical equations and formulae are derived for the quadratic logistic parameter estimation. In Section 3, the sample paths of capacity dimension, information dimension, and correlation dimension of fractal growth of Beijing city are employed as examples to show the model building process. In Section 4, the related questions are discussed, and finally, in Section 5, the discussion is concluded with outlining the key points of this work.

# 2 Models

## 2.1 Basic models and formulae

For the cities in the developed countries, the fractal dimension curves of urban form and growth can be modeled with logistic function. In contrast, for great majority of cities in China, the fractal dimension curves cannot be described with the ordinary logistic function. Generally speaking, the fractal dimension curves of Chinese cities can be described by means of quadratic logistic function (Chen, 2018). The quadratic logistic model of the fractal dimension curves of urban form is as follows

$$D(t) = \frac{D_{max}}{1+(D_{max}/D_{(0)}-1)e^{-(kt)^2}}, \tag{1}$$

in which $D(t)$ refers to the fractal dimension of urban form at time $t$, $D_{(0)}$ denotes the initial value of fractal dimension at time $t=0$, $D_{max}$ is the capacity value of fractal dimension, that is, the upper limit of fractal dimension, and $k$ is the initial growth rate of fractal dimension. Equation (1) can be changed into the following form

$$\ln(\frac{D_{max}}{D(t)}-1) = \ln(\frac{D_{max}}{D_{(0)}}-1)-(kt)^2 = \ln a - k^2 t^2. \tag{2}$$

Let

$$y_t = \frac{D_{max}}{D(t)}-1, \quad a = \frac{D_{max}}{D_{(0)}}-1. \tag{3}$$

Based on equation (3), equation (2) can be simplified as a linear relation as follows

$$\ln y_t = \ln a - k^2 t^2. \tag{4}$$

Using linear regression analysis, we can estimate the values of $a$ and $k$, where $a$ is the estimated value of $D_{max}/D_{(0)}-1$. Thus, in terms of equation (3), the initial value of fractal dimension can be given by the following formula:

$$\hat{D}_{(0)} = \frac{D_{max}}{a+1}, \tag{5}$$

which differs slightly from the observed value of the fractal dimension at time $t=0$. In above formula, the symbol "^" means that the value is estimated result. Once the capacity value of fractal dimension, $D_{max}$, is determined, a simple linear regressive analysis can be performed to estimate the values of $a$ and $k$. So, the key step is to estimate the value of the capacity parameter, $D_{max}$.

An effective approach to estimating the capacity parameter value of the quadratic logistic model is nonlinear regression analysis. Differentiating fractal dimension $D(t)$ in equation (1) with respect to time $t$ yields a growth speed equation as below:

$$\frac{dD(t)}{dt} = 2k^2 t D(t)[1 - \frac{D(t)}{D_{max}}], \qquad (6)$$

which represents the growth velocity of fractal dimension of urban form. Discretizing equation (6) yields a difference equation as follows

$$\frac{\Delta D(t)}{\Delta t} = \frac{D(t+\Delta t) - D(t)}{\Delta t} = 2k^2 t D(t) - \frac{2k^2 t}{D_{max}} D(t)^2 = btD(t) - ctD(t)^2. \qquad (7)$$

in which the parameters are $b=2k^2$ and $c=2k^2/D_{max}$. Thus we have a formula as follows

$$\hat{D}_{max} = \frac{2k^2}{c} = \frac{b}{c}, \qquad (8)$$

which suggests that the capacity value, $D_{max}$, can be estimated by $k$ and $c$ values. It can be seen that, based on equation (7), two parameters can be evaluated, one is the initial growth rate, $k$, and the other is the capacity parameter, $D_{max}$.

Next, we have two approaches to estimating the values of the remaining parameter(s). One is formula averaging method, and the other, is regression analysis method. Since the values of the initial growth rate, $k$, and the capacity parameter, $D_{max}$, have been estimated, the theoretical initial value of fractal dimension can be estimated by formula averaging method. Substituting equation (8) into equation (1) yields a series of initial values, which can be formulated as

$$\hat{D}_{(0)}(t_i) = \frac{\hat{D}_{max}}{1 + (\frac{\hat{D}_{max}}{D(t_i)} - 1)e^{(\hat{k}t_i)^2}}, \qquad (9)$$

where $i=1,2,3,\ldots,n$ denotes the number of sampling time. The initial value of fractal dimension as a parameter can be estimated by arithmetic mean, namely

$$\hat{D}_{(0)} = \frac{1}{n}\sum_{i=1}^{n}\hat{D}_{(0)}(t_i) = \frac{1}{n}\sum_{i=1}^{n}\frac{\hat{D}_{max}}{1 + (\frac{\hat{D}_{max}}{D(t)} - 1)e^{(\hat{k}t_i)^2}}. \qquad (10)$$

So far, we have worked out the estimated values of all parameters in the quadratic logistic model.

Another method is to use linear regression analysis to estimate the values of the initial growth rate and the initial value of fractal dimension, that is, $k$ and $D_{(0)}$. Although the $k$ value has been estimated by means of equation (7), it can be abandoned. Substituting the estimated value of the value of $D_{max}$ based on equation (7) into equation (2), we can make a linear regression analysis. The independent variable is $t^2$, and the dependent variable is $\ln y_t$. By the linear regression analysis, we can figure out the values of $a$ and $k$, where $a$ is the estimated value of $D_{max}/D_{(0)}-1$. The value of $k$ given by equation (2) is different slightly from the result from equation (7). Thus the initial value of fractal dimension can be estimated by means of equation (5). The estimated value of $D_{(0)}$ based on equation (5) is different to some extent from that based on equation (10). Substituting all the parameter values into equation (1), we have an empirical model of quadratic logistic growth, by which we can make city growth prediction analysis or stage division analysis.

**2.2 Nonlinear regressive equations and related formulae for regular sampling**

If the fractal dimension sampling points are regularly distributed, for example, the fractal dimension value is estimated once a year, or every other year, or every two years, we will have $\Delta t$=constant, and $\Delta D(t)= D(t+\Delta t)-D(t)$. Thus, equation (7) can be converted into the following expression

$$D(t+\Delta t) - D(t) = 2k^2 \Delta t (tD(t)) - \frac{2k^2 \Delta t}{D_{max}} (tD(t)^2) . \tag{11}$$

So, the logistic model parameters can be estimated with the following equation

$$D(t+\Delta t) = D(t) + 2k^2 \Delta t (tD(t)) - \frac{2k^2 \Delta t}{D_{max}} (tD(t)^2) = D(t) + b'(tD(t)) - c'(tD(t)^2) . \tag{12}$$

which is actually a nonlinear autoregressive equation. In equation (12), the regressive coefficients are $b'=2k^2\Delta t$ and $c'=2k^2\Delta t/D_{max}$, from which we can estimate the values of $k$ and $D_{max}$. The formulae are as below:

$$\hat{k}^2 = \frac{b'}{2\Delta t} , \tag{13}$$

$$\hat{D}_{max} = \frac{2\hat{k}^2 \Delta t}{c'} = \frac{b'}{c'} . \tag{14}$$

Then, we have two approaches to estimating the values of the remaining parameter(s): formula averaging method and regression analysis method. Base on equations (9) and (10), the formula averaging method can be used to estimate the value of parameter, $D_0$. Base on equations (2), (3), (4), and (5), the regression analysis method can be utilized to estimate the values of parameter $D_{(0)}$ and $k$, where the $k$ value belongs to the second estimation.

**2.3 Nonlinear regressive equations and related formulae for continuous sampling**

If the fractal dimension sampling points are continuously distributed, that is to say, fractal dimension values were calculated year by year, we will have $\Delta t=1$, and $\Delta D(t)= D(t+1) - D(t)$. Note that the so-called continuity here is not a mathematical continuity, but an empirical continuity. The continuity of mathematics leads to differentiation rather than difference. Thus, equation (11) can be converted into the following expression

$$D(t+1) - D(t) = 2k^2 t D(t) - \frac{2k^2}{D_{max}} t D(t)^2. \tag{15}$$

In this case, the logistic model parameters can be estimated by regressive analysis based on the following equation

$$D(t+1) = D(t) + 2k^2 t D(t) - \frac{2k^2}{D_{max}} t D(t)^2 = D(t) + b'' t D(t) - c'' t D(t)^2. \tag{16}$$

where the regressive coefficients are $b''=2k^2$ and $c''=2k^2/D_{max}$. Equation (16) is a clear nonlinear autoregressive equation. It suggests that equation (7) is essentially a nonlinear autoregressive equation. By using equation (16), we can estimate the values of $k$ and $D_{max}$. The formulae of estimating the values of $k$ and $D_{max}$ are as follows

$$\hat{k}^2 = b/2, \tag{17}$$

$$\hat{D}_{max} = \frac{2\hat{k}^2}{c''} = \frac{b''}{c''}, \tag{18}$$

which are special cases of equations (13) and (14). By comparison, we can see the relationships, similarities, and differences between equation (8), equation (14), and equation (18).

Then, we have two methods to estimate the values of the remaining parameter(s). Similar to the previous two situations, one is the formula averaging method, and the other is the regression analysis method. The formula averaging method is as follows. The initial value of fractal dimension can be estimated by

$$\hat{D}_{(0)}(t) = \frac{D_{max}}{1+(\frac{D_{max}}{D(t)}-1)e^{kt^2}}. \tag{19}$$

The formula of the average value is

$$\hat{D}_{(0)} = \frac{1}{n}\sum_{t=1}^{n}\hat{D}_{(0)}(t) = \frac{1}{n}\sum_{t=1}^{n}\frac{\hat{D}_{max}}{1+(\frac{\hat{D}_{max}}{D(t)}-1)e^{\hat{k}t^2}}. \tag{20}$$

As for the regression analysis method, substituting the value given by equation (18) into equation (2), we can make a regression analysis to estimate the values of $D_{(0)}$ and $k$, where the $k$ value is the result from the second estimation.

Table 1 Parameter estimation methods of logistic model of fractal dimension curve of urban growth based on combination of different equations

| Sampling | Approach | Equation combination |
|---|---|---|
| **Random sampling** | Approach 1 | Equations (7) and (8); Equations (9) and (10) |
| | Approach 2 | Equations (7) and (8); Equations (2), (3), (4), and (5) |
| **Regular sampling,** | Approach 1 | Equations (12), (13), and (14); Equations (9) and (10) |
| | Approach 2 | Equations (12), (13), and (14); Equations (2), (3), (4), and (5) |
| **Continuous sampling** | Approach 1 | Equations (16), (17), and (18); Equations (19) and (20) |
| | Approach 2 | Equations (16), (17), and (18); Equations (2), (3), (4), and (5) |

**Note**: In approach 1 for continuous sampling sequence, equations (19) and (20) can be replaced by equations (9) and (10), equivalently.

## 2.4 Method summarization of parameter estimation

After deriving a mathematical model, corresponding parameter estimation methods can be designed. Based on the equations and formulae derived above, three sets of approaches can be designed to estimate the parameter values of logistic models of fractal dimension growth (Table 1). The first set of methods is for random sampling sequence, the second set of methods is for regular sampling sequence, and the third set of methods is for continuous sampling sequence. The method for random sequences can be used for both regular and continuous sequences, while the method for regular sequences can be used for continuous sequences. But the opposite is not true.

1. Two approaches for model parameter estimation based on random sampling sequence of fractal

dimension. (1) Nonlinear bivariate auto-regression + formulae. The first approach is to make use of equation (7), (8), (9), and (10). By means of equations (7) and (8), a nonlinear auto-regression analysis can be made to estimate the values of capacity parameters $D_{max}$ and inherent growth rate $k$. Then, by using equations (9) and (10), the initial value of fractal dimension, $D_{(0)}$, can be estimated. Substituting the estimated results into equation (1) yields a logistic model of fractal dimension curve based on random sampling. (2) Nonlinear bivariate auto-regression + linear regression + formulae. The second approach is to make use of equations (2), (3), (4), (5), (7), and (8). Using equations (7) and (8) to estimate the capacity parameter $D_{max}$, and estimate the initial growth rate $k$ preliminarily. Then, using equations (2), (3), (4) and (5) to estimate the parameter $D_{(0)}$, and estimate the initial growth rate $k$ once again. Substituting the estimated results into equation (1) yields the second logistic model of fractal dimension curve based on random sampling.

2. Two approaches for model parameter estimation based on regular sampling sequence of fractal dimension. (1) Nonlinear bivariate auto-regression + formulae. The first approach is to make use of equations (9), (10), (12), (13), and (14). By means of equations (12), (13), and (14), a nonlinear auto-regression analysis can be made to estimate the values of capacity parameters $D_{max}$ and inherent growth rate $k$. Then, by using equations (9) and (10), the initial value of fractal dimension, $D_{(0)}$, can be estimated. Substituting the estimated results into equation (1) yields a logistic model of fractal dimension curve based on regular sampling. (2) Nonlinear bivariate auto-regression + linear regression + formulae. The second approach is to make use of equations (2), (3), (4), (5), (12), (13), and (14). Using equations (12), (13), and (14) to estimate the capacity parameter $D_{max}$, and estimate the initial growth rate $k$ preliminarily. Then, using equations (2), (3), (4) and (5) to estimate the parameter $D_{(0)}$, and estimate the initial growth rate $k$ once again. Substituting the estimated results into equation (1) yields the second logistic model of fractal dimension curve based on regular sampling.

3. Two approaches for model parameter estimation based on continuous sampling sequence of fractal dimension. (1) Nonlinear bivariate auto-regression + formulae. The first approach is to make use of equations (16), (17), (18), (19), and (20). By means of equations (16), (17), and (18), a nonlinear auto-regression analysis can be made to estimate the values of capacity parameters $D_{max}$ and inherent growth rate $k$. Then, by using equations (19) and (20), the initial value of fractal dimension, $D_{(0)}$, can be estimated. Equations (19) and (20) are the special cases of equations (9) and (10). So the former can be replaced with the latter. Substituting the estimated results into equation (1) yields a logistic model of fractal dimension curve based on continuous sampling. (2) Nonlinear bivariate auto-regression + linear regression + formulae. The second approach is to make use of equations (2), (3), (4), (5), (16), (17), and (18). Using equations (16), (17), and (18) to estimate the capacity parameter $D_{max}$, and estimate the initial growth rate $k$ preliminarily. Then, using equations (2), (3), (4) and (5) to estimate the parameter $D_{(0)}$, and estimate the initial growth rate $k$ once again. Substituting the estimated results into equation (1) yields the second logistic model of fractal dimension curve based on continuous sampling.

The above approaches are theoretically equivalent to one another, but empirically they may result in different parameter values. However, there is no significant difference in the same parameter's values based on different methods. All approaches involve regression and auto-regression, with nonlinear auto-regression conducted through bivariate linear regression (Table 2).

Table 2 Independent variables and dependent variables in four sets of regression analysis

| Type | Argument | Lag variable | Parameter | Equation |
| --- | --- | --- | --- | --- |
| **Basic model** | $t$ | $\ln(D_{max}/D(t)-1)$ | $a=\ln(D_{max}/D_0-1)$, $k$ | Equations (2) |
| **Random sequence** | $D(t)$, $D(t)^2$ | $\Delta D(t)/\Delta t$ | $b=k$, $c=k/D_{max}$ | Equations (7) |
| **Regular sequence** | $D(t)$, $D(t)^2$ | $D(t+\Delta t)$ | $b=1+k\Delta t$, $c= k\Delta t/D_{max}$ | Equations (12) |
| **Continuous sequence** | $D(t)$, $D(t)^2$ | $D(t+1)$ | $b=1+k$, $c= k/D_{max}$ | Equations (16) |

# 3 Empirical analysis

## 3.1 Study area and area

The effectiveness of the method can be reflected through case analysis. The city of Beijing, the capital of China, is taken as example to illustrate the newly developed parameter estimation method. Three types of multifractal parameters are taken into account, that is, capacity dimension $D_0$, information dimension $D_1$, and correlation dimension $D_2$. All these parameters belong to generalized correlation dimension set, which are defined at the macro level. The capacity dimension reflects space-filling degree, the information dimension reflect spatial uniformity degree, and correlation dimension reflects spatial dependence degree. The calculation results of fractal parameters of urban morphology in Beijing over the past 33 years are as follows, with a time span of data from 1985 to 2017 (Table 3).

Table 3 The capacity dimension, information dimension, and correlation dimension of Beijing's urban form (1985–2017)

| Year | $D_0$ | $D_1$ | $D_2$ | Year | $D_0$ | $D_1$ | $D_2$ |
|------|-------|-------|-------|------|-------|-------|-------|
| 1985 | 1.8115 | 1.7340 | 1.7147 | 2002 | 1.8964 | 1.8411 | 1.8258 |
| 1986 | 1.8176 | 1.7405 | 1.7218 | 2003 | 1.9030 | 1.8501 | 1.8350 |
| 1987 | 1.8284 | 1.7563 | 1.7394 | 2004 | 1.9089 | 1.8586 | 1.8438 |
| 1988 | 1.8311 | 1.7594 | 1.7428 | 2005 | 1.9131 | 1.8648 | 1.8503 |
| 1989 | 1.8342 | 1.7636 | 1.7472 | 2006 | 1.9182 | 1.8717 | 1.8574 |
| 1990 | 1.8373 | 1.7680 | 1.7519 | 2007 | 1.9232 | 1.8788 | 1.8646 |
| 1991 | 1.8398 | 1.7707 | 1.7546 | 2008 | 1.9271 | 1.8846 | 1.8708 |
| 1992 | 1.8423 | 1.7738 | 1.7578 | 2009 | 1.9302 | 1.8900 | 1.8767 |
| 1993 | 1.8463 | 1.7786 | 1.7628 | 2010 | 1.9339 | 1.8963 | 1.8837 |
| 1994 | 1.8500 | 1.7834 | 1.7679 | 2011 | 1.9369 | 1.9012 | 1.8891 |
| 1995 | 1.8542 | 1.7884 | 1.7729 | 2012 | 1.9399 | 1.9057 | 1.8941 |
| 1996 | 1.8575 | 1.7928 | 1.7775 | 2013 | 1.9429 | 1.9106 | 1.8995 |
| 1997 | 1.8612 | 1.7976 | 1.7825 | 2014 | 1.9466 | 1.9164 | 1.9058 |
| 1998 | 1.8666 | 1.8042 | 1.7892 | 2015 | 1.9501 | 1.9221 | 1.9120 |
| 1999 | 1.8715 | 1.8104 | 1.7955 | 2016 | 1.9523 | 1.9247 | 1.9148 |
| 2000 | 1.8797 | 1.8201 | 1.8050 | 2017 | 1.9541 | 1.9302 | 1.9213 |
| 2001 | 1.8905 | 1.8334 | 1.8180 |      |        |        |        |

**Note**: The original data sources are satellite remote sensing images with a spatial resolution of 30 m, including TM data from the US Land Resources Satellite and CCD data from domestic environmental satellites. The interpretation method is supervised classification, with a classification accuracy of over 86%.

## 3.2 Calculation results

The first procedure is to organize data. Time is defined as $t=n-n_0=n-1985$, where $n$ denotes year, and $n_0=1985$ refers to initial year. Independent variables are $tD_q(t)$ and $tD_q(t)^2$, and dependent variable is $\Delta D_q(t)/\Delta t=(D_q(t+\Delta t)-D_q(t))/\Delta t$, where $q=0, 1, 2$ is order of moment for multifractal dimension, and $\Delta t$ represents time difference. Since $\Delta t=1$, we have $\Delta D_q(t)/\Delta t=\Delta D_q(t)$ (Table 4). Due to the continuous and regularly spaced fractal dimension values, we can use the third set of equations, i.e., equations (15) to (20), to estimate the model parameter values, and we can also use the second set of equations, i.e., equations (11) to (14), to estimate the model parameter values. For generality, the first set of equations, i.e., equations (6) to (10), are employed to estimate the model parameter values.

Table 4 The arranged results of multifractal dimension data for nonlinear regression analysis

| Time | Capacity dimension $D_0$ | | | Information dimension $D_1$ | | | Correlation dimension $D_2$ | | |
|---|---|---|---|---|---|---|---|---|---|
| (t) | $tD_0(t)$ | $tD_0(t)^2$ | $\Delta D_0(t)$ | $tD_1(t)$ | $tD_1(t)^2$ | $\Delta D_1(t)$ | $tD_2(t)$ | $tD_2(t)^2$ | $\Delta D_2(t)$ |
| 0  | 0.0000  | 0.0000   | 0.0062 | 0.0000  | 0.0000   | 0.0065 | 0.0000  | 0.0000   | 0.0071 |
| 1  | 1.8176  | 3.3038   | 0.0108 | 1.7405  | 3.0294   | 0.0158 | 1.7218  | 2.9646   | 0.0176 |
| 2  | 3.6568  | 6.6861   | 0.0027 | 3.5126  | 6.1693   | 0.0031 | 3.4789  | 6.0512   | 0.0033 |
| 3  | 5.4933  | 10.0587  | 0.0031 | 5.2783  | 9.2869   | 0.0041 | 5.2283  | 9.1117   | 0.0044 |
| 4  | 7.3369  | 13.4574  | 0.0031 | 7.0542  | 12.4406  | 0.0044 | 6.9887  | 12.2103  | 0.0047 |
| 5  | 9.1867  | 16.8792  | 0.0025 | 8.8399  | 15.6289  | 0.0027 | 8.7594  | 15.3453  | 0.0027 |
| 6  | 11.0388 | 20.3092  | 0.0025 | 10.6242 | 18.8122  | 0.0031 | 10.5276 | 18.4719  | 0.0032 |
| 7  | 12.8961 | 23.7583  | 0.0040 | 12.4165 | 22.0241  | 0.0048 | 12.3045 | 21.6287  | 0.0050 |
| 8  | 14.7703 | 27.2701  | 0.0038 | 14.2290 | 25.3080  | 0.0048 | 14.1027 | 24.8606  | 0.0051 |
| 9  | 16.6504 | 30.8042  | 0.0041 | 16.0507 | 28.6252  | 0.0050 | 15.9110 | 28.1288  | 0.0050 |
| 10 | 18.5417 | 34.3793  | 0.0033 | 17.8837 | 31.9827  | 0.0044 | 17.7290 | 31.4319  | 0.0046 |
| 11 | 20.4325 | 37.9533  | 0.0037 | 19.7208 | 35.3555  | 0.0048 | 19.5527 | 34.7551  | 0.0050 |
| 12 | 22.3347 | 41.5700  | 0.0053 | 21.5709 | 38.7754  | 0.0066 | 21.3899 | 38.1272  | 0.0067 |
| 13 | 24.2653 | 45.2927  | 0.0049 | 23.4540 | 42.3147  | 0.0062 | 23.2599 | 41.6170  | 0.0063 |
| 14 | 26.2007 | 49.0341  | 0.0082 | 25.3456 | 45.8857  | 0.0097 | 25.1367 | 45.1324  | 0.0095 |
| 15 | 28.1951 | 52.9974  | 0.0108 | 27.3021 | 49.6937  | 0.0133 | 27.0748 | 48.8698  | 0.0130 |
| 16 | 30.2483 | 57.1848  | 0.0059 | 29.3347 | 53.7827  | 0.0077 | 29.0877 | 52.8808  | 0.0079 |
| 17 | 32.2393 | 61.1394  | 0.0066 | 31.2983 | 57.6225  | 0.0091 | 31.0394 | 56.6731  | 0.0092 |
| 18 | 34.2536 | 65.1839  | 0.0059 | 33.3023 | 61.6134  | 0.0085 | 33.0303 | 60.6111  | 0.0088 |
| 19 | 36.2690 | 69.2336  | 0.0042 | 35.3134 | 65.6335  | 0.0062 | 35.0331 | 64.5958  | 0.0065 |
| 20 | 38.2615 | 73.1969  | 0.0051 | 37.2966 | 69.5516  | 0.0069 | 37.0068 | 68.4751  | 0.0071 |
| 21 | 40.2820 | 77.2685  | 0.0050 | 39.3065 | 73.5714  | 0.0070 | 39.0055 | 72.4491  | 0.0072 |
| 22 | 42.3108 | 81.3729  | 0.0038 | 41.3328 | 77.6544  | 0.0058 | 41.0217 | 76.4902  | 0.0062 |
| 23 | 44.3223 | 85.4114  | 0.0032 | 43.3454 | 81.6882  | 0.0054 | 43.0293 | 80.5008  | 0.0059 |
| 24 | 46.3254 | 89.4185  | 0.0036 | 45.3596 | 85.7288  | 0.0063 | 45.0406 | 84.5274  | 0.0070 |
| 25 | 48.3463 | 93.4948  | 0.0031 | 47.4075 | 89.8987  | 0.0049 | 47.0917 | 88.7051  | 0.0054 |
| 26 | 50.3598 | 97.5426  | 0.0030 | 49.4300 | 93.9741  | 0.0046 | 49.1160 | 92.7840  | 0.0051 |
| 27 | 52.3767 | 101.6045 | 0.0030 | 51.4549 | 98.0594  | 0.0049 | 51.1419 | 96.8700  | 0.0054 |
| 28 | 54.4007 | 105.6942 | 0.0038 | 53.4971 | 102.2122 | 0.0058 | 53.1871 | 101.0310 | 0.0063 |
| 29 | 56.4527 | 109.8933 | 0.0034 | 55.5766 | 106.5090 | 0.0056 | 55.2688 | 105.3326 | 0.0062 |
| 30 | 58.5016 | 114.0814 | 0.0022 | 57.6621 | 110.8307 | 0.0026 | 57.3610 | 109.6762 | 0.0028 |
| 31 | 60.5200 | 118.1508 | 0.0019 | 59.6651 | 114.8364 | 0.0055 | 59.3584 | 113.6588 | 0.0066 |
| 32 | 62.5318 | 122.1947 |        | 61.7665 | 119.2219 |        | 61.4828 | 118.1294 |        |

**Note**: The last row of data cannot be used in parameter estimation process. Due to $\Delta t =1$, $\Delta D_q(t)/\Delta t = \Delta D_q(t)$.

The second procedure is to make nonlinear regression analysis. The key mathematical expression is equation (7). The related formulae are equations (9) and (10). By conducting nonlinear regression analysis, we can obtain the estimated values of the capacity parameter value of the quadratic logistic models; the initial growth rate can also be estimated; by means of equations (9) and (10), we can estimate the initial value of fractal dimension (Table 5). Let make models one by one.

(1) Capacity dimension. For capacity dimension $D_0$, two dependent variables are $tD_0(t)$ and $tD_0(t)^2$, the dependent variable is $\Delta D_0(t)/\Delta t = \Delta D_0(t)$, in which $\Delta t = 1$. The constant term is zero. By least squares calculation, we have a model as

$$\frac{\Delta \hat{D}_0(t)}{\Delta t} = \Delta \hat{D}_0(t) = 0.005151 t D_0(t) - 0.002623 t D_0(t)^2. \tag{21}$$

The goodness of fit is about $R^2=0.7611$. According to equation (8), the capacity parameter of capacity dimension is $D_{\max}^{(0)} = 0.005151/0.002623 = 1.9636$. We can also estimate the inherent growth rate of capacity dimension, that is, $k=(0.005151/2)^{0.5} = 0.0508$. Using equations (9) and (10), we can estimate the initial fractal parameters, that is, $D_0(0)= 1.8253$, thus we have $a= D_{\max}/D_0(0)-1=0.0758$. Thus we have a quadratic logistic growth model for capacity dimension of Beijing's urban form as follows

$$\hat{D}_0(t) = \frac{1.9636}{1+(1.9636/1.8253-1)e^{-(0.0508t)^2}} = \frac{1.9636}{1+0.0758 e^{-(0.0508t)^2}}. \tag{22}$$

The predicted values generated by this model match the observed data of capacity dimension very well (Figure 1(a)).

(2) Information dimension. For information dimension $D_1$, two dependent variables are $tD_1(t)$ and $tD_1(t)^2$, the dependent variable is $\Delta D_1(t)/\Delta t = \Delta D_1(t)$. The constant term is still set as zero. Using least squares method, we obtain a model as below

$$\frac{\Delta \hat{D}_1(t)}{\Delta t} = \Delta \hat{D}_1(t) = 0.004348 t D_1(t) - 0.002227 t D_1(t)^2. \tag{23}$$

The goodness of fit is about $R^2=0.7895$. The capacity parameter of information dimension is $D_{\max}^{(1)} = 0.004348/0.002227 = 1.9523$. We can also estimate the inherent growth rate of information dimension, that is, $k=(0.004348/2)^{0.5} = 0.0466$. Using equations (9) and (10), we can estimate the initial fractal parameters, that is, $D_1(0)= 1.7516$, thus we have $a= D_{\max}/D_0(0)-1=0.1146$. Thus we have a quadratic logistic growth model for information dimension of Beijing's urban form as below

$$\hat{D}_1(t) = \frac{1.9523}{1+(1.9523/1.7516-1)e^{-(0.0466t)^2}} = \frac{1.9523}{1+0.1146 e^{-(0.0466t)^2}}. \tag{24}$$

The predicted values produced by this model match the observed data of the information dimension very well (Figure 1(b)).

(3) Correlation dimension. For correlation dimension $D_2$, two dependent variables are $tD_2(t)$ and $tD_2(t)^2$, the dependent variable is $\Delta D_2(t)/\Delta t = \Delta D_2(t)$. The constant term is still set as zero. Through least squares computation, we get a model as follows

$$\frac{\Delta \hat{D}_2(t)}{\Delta t} = \Delta \hat{D}_2(t) = 0.004104 t D_2(t) - 0.002107 t D_2(t)^2. \tag{25}$$

The goodness of fit is about $R^2=0.7726$. The capacity parameter of correlation dimension is $D_{max}^{(2)}$ = 0.004104/0.002107=1.9483. We can also estimate the inherent growth rate of correlation dimension, that is, $k=(0.004104/2)^{0.5} = 0.0453$. Using equations (9) and (10), we can estimate the initial fractal parameters, that is, $D_2(0)= 1.7350$, thus we have $a= D_{max}/D_0(0)-1=0.1229$. Thus we have a quadratic logistic growth model for correlation dimension of Beijing's urban form as follows

$$\hat{D}_2(t) = \frac{1.9483}{1+(1.9483/1.7350-1)e^{-(0.0453t)^2}} = \frac{1.9483}{1+0.1229 e^{-(0.0453t)^2}}. \tag{26}$$

The predicted values yielded by this model match the observed data of correlation dimension very well (Figure 1(c)).

Generally speaking, if the following conditions are met, the above model is sufficient. Firstly, there is a large time span, a large number of sample points, or a sufficiently long sample path. Secondly, the quality of fractal data is good. Thirdly, there is a determinate pattern of urban growth. In this case, equations (22), (24), and (26) can be used to predict urban growth and make spatial dynamical analysis of urban morphology of Beijing.

Table 5 Coefficient estimation results based on nonlinear regression and related statistics

| Parameter | Capacity dimension | | Information dimension | | Correlation dimension | |
|---|---|---|---|---|---|---|
| Statistic | Coefficient | P-value | Coefficient | P-value | Coefficient | P-value |
| $b=2k^2$ | 0.005151 | 2.4014E-06 | 0.004348 | 8.6519E-06 | 0.004104 | 3.7124E-05 |
| $c=2k^2/D_{max}$ | 0.002623 | 3.3016E-06 | 0.002227 | 1.4452E-05 | 0.002107 | 6.2602E-05 |
| $D_{max}=b/c$ | 1.9636 | | 1.9523 | | 1.9483 | |
| $k=(b/2)^{1/2}$ | 0.0508 | | 0.0466 | | 0.0453 | |
| $D_{(0)}$ | 1.8253 | | 1.7516 | | 1.7350 | |
| $a= D_{max}/D_{(0)}-1$ | 0.0758 | | 0.1146 | | 0.1229 | |
| $R^2$ | 0.7611 | | 0.7895 | | 0.7726 | |

**Note**: The initial growth rates, that is, $k$ values, are estimated by nonlinear regression analysis based on equation (7).

The third procedure is to make linear regression analysis. The mathematical expression is equation (2) or equations (3) and (4). Sometimes, the initial parameter estimation of $D_q$ based on equations (9) and (10) may not be satisfying due to significant random perturbations in the fractal dimension calculation values. Concretely speaking, the estimated value of $D_q(0)$ and thus the value $a=D_{max}/D_q(0)-1$ is significantly biased. In this case, we had better re-estimate the values of $k$ and $a=D_{max}/D_q(0)-1$. Substituting the estimated values of $D_{max}$ into equation (2) or equation (3), we can making linear regression analysis by means of equation (2) or equation (4) to yield the estimated values of ln$a$ and $k^2$. Using formula $a=D_{max}/D_{(0)}-1$, we can estimate the value of $D_{(0)}$, which represents the predicted initial value of fractal dimension (Table 6). Generally speaking, a predicted value differs to some extent from the corresponding observed value. There is a slight difference between the second estimated inherent growth rate and the first estimated inherent growth rate, i.e., $k$ value.

**Table 6 The estimation results of quadratic logistic model parameters and related statistics**

| Parameter | Capacity dimension $D_0$ | Information dimension $D_1$ | Correlation dimension $D_2$ |
|---|---|---|---|
| $D_{max}$ | 1.9636 | 1.9523 | 1.9483 |
| $k$ | 0.0512 | 0.0467 | 0.0453 |
| $a=D_{max}/D_{(0)}-1$ | 0.0771 | 0.1148 | 0.1229 |
| $D_{(0)}$ | 1.8231 | 1.7513 | 1.7352 |
| $R^2$ | 0.9972 | 0.9975 | 0.9973 |

**Note**: The $k$ values are estimated by linear regression analysis based on equation (2). There are a slight differences between the two times of estimation results, but the differences are not significant.

The quadratic logistic models of fractal dimension curves can be built with newly estimated parameter values. For capacity dimension curve, the model is

$$\hat{D}_0(t) = \frac{1.9636}{1+(1.9636/1.8231-1)e^{-(0.0512t)^2}} = \frac{1.9636}{1+0.0771e^{-(0.0512t)^2}}. \quad (27)$$

The goodness of fit is about $R^2=0.9972$. The predicted values generated by equation (27) are generally consistent with the predicted values given by equation (22) (Figure 1(a)). For information dimension curve, the model is

$$\hat{D}_1(t) = \frac{1.9523}{1+(1.9523/1.7513-1)e^{-(0.0457t)^2}} = \frac{1.9523}{1+0.1148e^{-(0.0467t)^2}}. \tag{28}$$

The goodness of fit is about $R^2$=0.9975. The predicted values generated by equation (28) are much consistent with the predicted values given by equation (24) (Figure 1(b)). For correlation dimension curve, the model is

$$\hat{D}_2(t) = \frac{1.9483}{1+(1.9483/1.7352-1)e^{-(0.0453t)^2}} = \frac{1.9483}{1+0.1229e^{-(0.0453t)^2}}. \tag{29}$$

The goodness of fit is about $R^2$=0.9973. The predicted values generated by equation (29) are almost completely consistent with the predicted values given by equation (26) (Figure 1(b)) (Figure 1(c)).

For the case of Beijing city, two sets of models of fractal dimension curves are consistent with each other. That is to say, there is no significantly difference between the results based on the first method and the results based on the second method. This suggests that, on the one hand, the time series data quality of the multifractal parameters of Beijing's urban form is relatively high, and on the other hand, Beijing's urban growth follows the quadratic logistic model well. Using these models, we can predict urban development of Beijing, and make stage division analysis for the city.

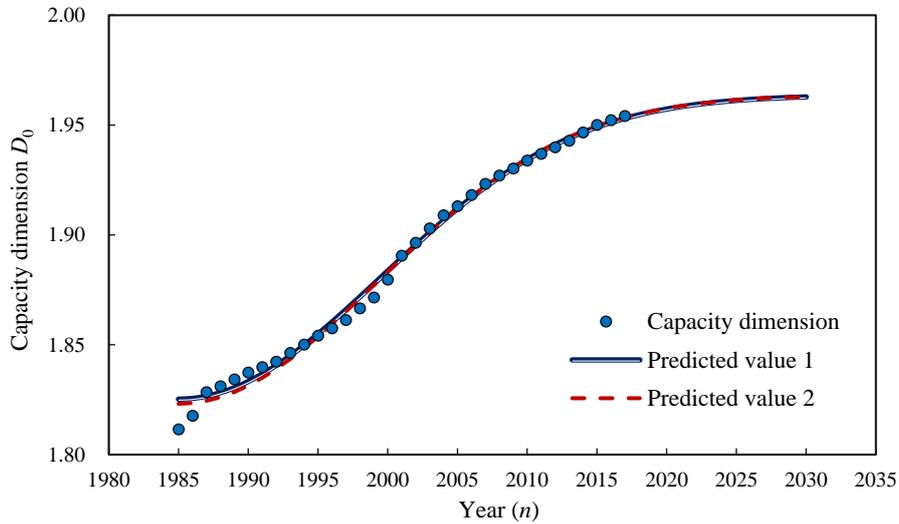

a. Capacity dimension $D_0$

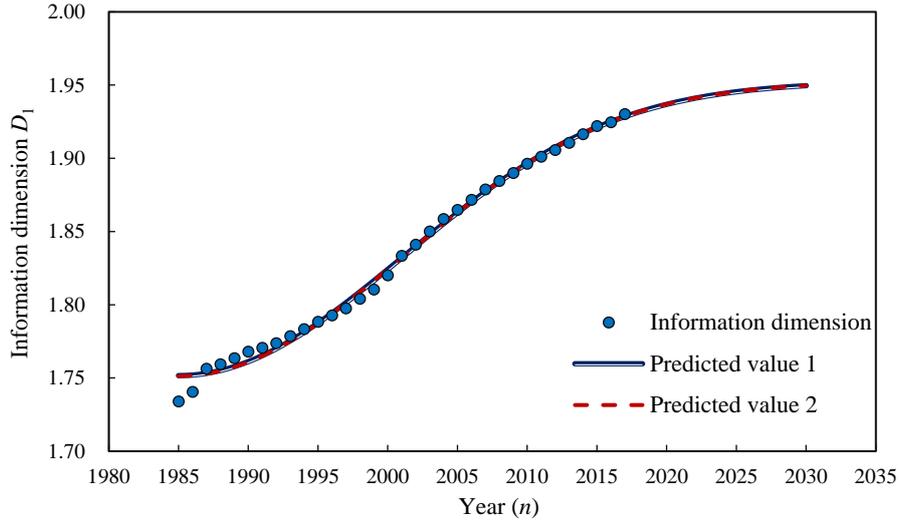

b. Information dimension $D_1$

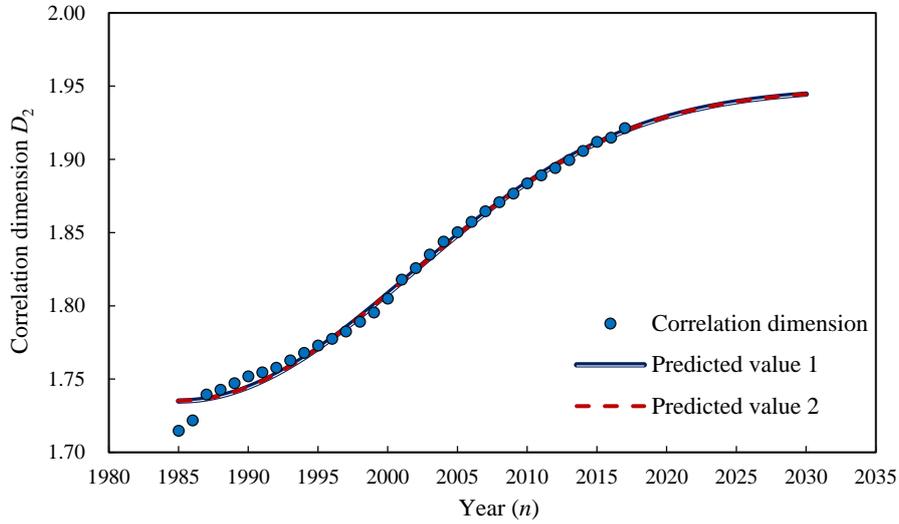

c. Correlation dimension $D_2$

**Figure 1. The observed values and predicted values of typical multifractal fractal dimension of urban growth of Beijing, China**

[**Note**: The first set of predicted values, "predicted value 1", are given by equations (22), (24), and (26), and the second set of predicted values, "predicted value 2", are produced by equations (27), (28), and (29). The observed values of the first and second years deviate slightly from the trend line. They look like two outliers. The reason may be due to the poor quality of early remote sensing images.]

# 4 Discussion

The calculation results show that the integrated framework of bivariate nonlinear regression analysis and ordinary linear regression analysis is one of effective approach for estimating the parameter values of quadratic logistic model of urban fractal dimension curves. Deriving the quadratic logistic growth model yields the growth speed equation. Discretizing the growth speed equation yields a bivariate nonlinear regression equation. Using the nonlinear regression equation, we can estimate the value of the capacity parameter. Substituting the estimated value of the capacity parameter into the quadratic logistic model and transforming the model into a quasilinear form, we can estimate the other parameter values by means of ordinary linear regression analysis. In this way, we can obtain all the estimated value of the quadratic logistic models (Figure 2).

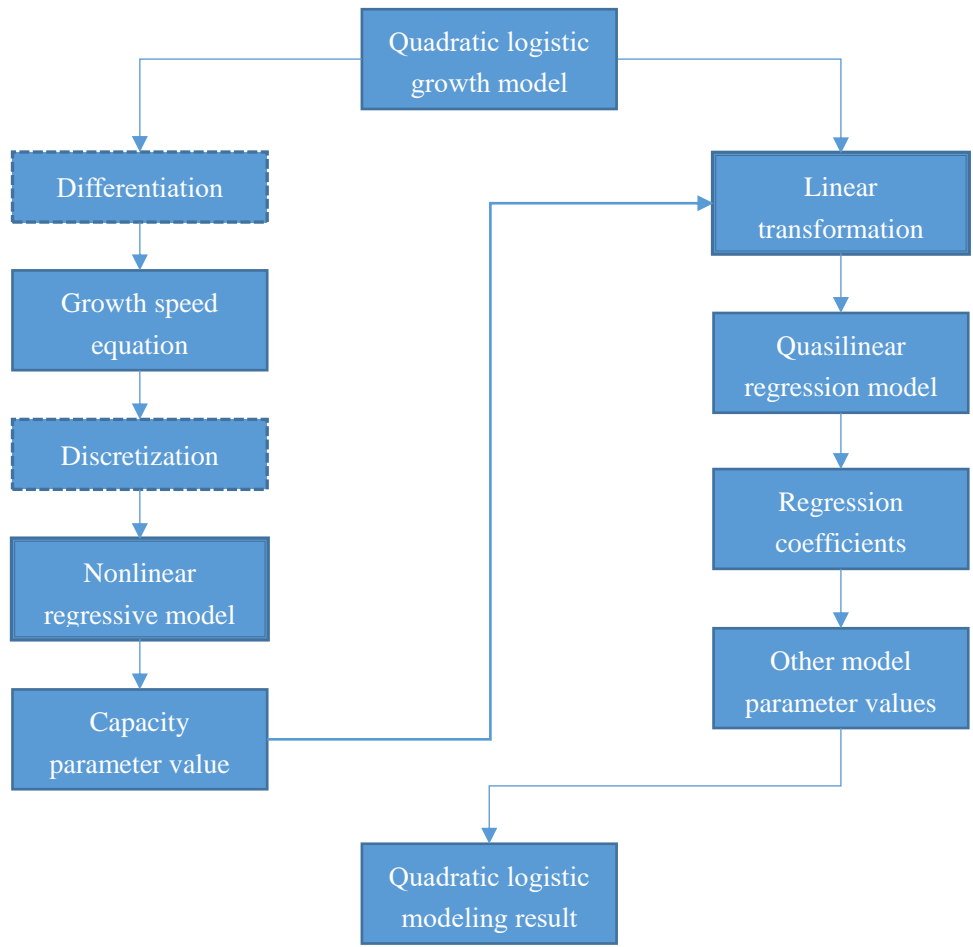

**Figure 2. Flow chart of parameter estimation for quadratic logistic model of fractal dimension curves of urban form**

[**Note**: There are two key steps in the parameter estimation process: (1) using the bivariate nonlinear regression equation to estimate the capacity parameter, $D_{max}$, of fractal dimension curve model; (2) using quasilinear regression model to estimate the rest parameter value of the quadratic logistic model, i.e., the initial value, $D_{(0)}$, and the initial growth rate of fractal dimension, $k$.]

The method of estimating parameter values of common logistic model of fractal urban growth is relative simple. Combining ordinary linear regression analysis of nonlinear autoregressive analysis, we can work out all the parameter value of logistic models of fractal dimension curves. The key model is a nonlinear autoregressive equation. In contrast, for the quadratic logistic modeling, the key model is a nonlinear equation, but not a typical nonlinear autoregressive model. The time variable participates in the nonlinear regression process. This not only increases the difficulty of model parameter estimation, but also reduces the stability of parameter estimation results. Comparing the similarities and differences between two types of models and their parameter estimation equations helps to understand and apply parameter estimation methods for modeling process based on S-shaped functions (Table 7). Moreover, similar to ordinary logistic model, the parameter values of quadratic logistic models can also be worked out by using brute force search method and curve-fitting method.

Table 7 A comparison between parameter estimation equations and formulae of ordinary logistic model and quadratic logistic model

| Model | Subtype | Ordinary logistic model | Quadratic logistic model |
|---|---|---|---|
| **Basic model** | Growth model | $D(t) = \dfrac{D_{max}}{1+(D_{max}/D_{(0)}-1)e^{-kt}}$ | $D(t) = \dfrac{D_{max}}{1+(D_{max}/D_{(0)}-1)e^{-(kt)^2}}$ |
| | Speed equation | $\dfrac{dD(t)}{dt} = kD(t)(1-\dfrac{D(t)}{D_{max}})$ | $\dfrac{dD(t)}{dt} = 2k^2 tD(t)[1-\dfrac{D(t)}{D_{max}}]$ |
| **Key model** | Nonlinear regression | $\dfrac{\Delta D(t)}{\Delta t} = bD(t)-cD(t)^2$ | $\dfrac{\Delta D(t)}{\Delta t} = btD(t)-ctD(t)^2$ |
| | Parameter formula | $\hat{D}_{max} = \dfrac{k}{c} = \dfrac{b}{c}$ | $\hat{D}_{max} = \dfrac{2k^2}{c} = \dfrac{b}{c}$ |
| | | $\hat{D}_{(0)} = \dfrac{1}{n}\sum_{i=1}^{n} \dfrac{\hat{D}_{max}}{1+(\dfrac{\hat{D}_{max}}{D(t)}-1)e^{\hat{k}t_i}}$ | $\hat{D}_{(0)} = \dfrac{1}{n}\sum_{i=1}^{n} \dfrac{\hat{D}_{max}}{1+(\dfrac{\hat{D}_{max}}{D(t)}-1)e^{(\hat{k}t_i)^2}}$ |
| **Linear model** | Ordinary regression | $\ln(\dfrac{D_{max}}{D(t)}-1) = \ln a - kt$ | $\ln(\dfrac{D_{max}}{D(t)}-1) = \ln a - k^2 t^2$ |
| | Parameter formula | $a = \dfrac{D_{max}}{D_{(0)}} - 1$ | $a = \dfrac{D_{max}}{D_{(0)}} - 1$ |

Compared with previous works, the novelty of this study rests with four aspects. (1) First, a set of equations and formulae for parameter estimation of quadratic logistic models of fractal dimension curves are derived by mathematical reasoning. (2) The approaches of the related parameter estimation are outlined. (3) A case analysis is presented to demonstrate the newly developed methods. The main shortcomings of this method lies in two aspects. Frist, the characteristics of observation data have a significant impact on the practicality and application effectiveness of the method. The effective application of the method developed in this work requires three prerequisites. (1) The data quality is good. (2) Sample path is long. (3) Growth of the system is stable. It goes without saying that inaccurate data inevitably leads to significant parameter estimation bias. The data quality is high, but short sample path also results in biased estimation of the model parameters. Sometimes, the system growth is unstable, and the fractal dimension curve exhibits significant disturbances, even jumping changes, there may also be issues with model parameter estimation by using the newly developed method. Second, the parameter estimation methods developed in this work are only suitable for the quadratic logistic model. The models of fractal dimension curves of urban form include the ordinary logistic function, quadratic logistic function, fractional logistic function, Boltzmann equation, and so on. Only the quadratic logistic model is take into consideration in this research. The methods of estimating the parameter values of ordinary logistic model has been advanced. Further development is needed for parameter estimation methods of fractional logistic model in the future.

# 5 Conclusions

In the natural and social fields, quadratic logistic growth is a common phenomenon. Therefore, using quadratic logistic functions to establish mathematical models is necessary. However, estimating model parameters is not an easy task. This paper proposes an effective method for parameter estimation in quadratic logistic models. The main conclusions can be reached as follows.

*First, an effective method of estimating parameter values of quadratic logistic models is to integrate bivariate nonlinear regression and ordinary linear regression into a framework*. A growth speed equation of fractal dimension of urban form can be derived from quadratic logistic function. Discretizing the growth speed equation yields a bivariate nonlinear regression equation with time variable. Using the bivariate nonlinear regression equation, we can estimate the capacity parameter

value of the quadratic logistic model. Then, transforming the quadratic logistic function into a quasilinear relation, we can estimate the initial value and inherent growth rate of fractal dimension. Thus, we obtain the whole parameter values for the quadratic logistic model of a fractal dimension curve. *Second, there are similarities and differences between the method of estimating the quadratic logistic model parameters and that of estimating the ordinary logistic model parameters*. The macro mathematical structure of quadratic logistic function is similar to that of common logistic function. This determines that there are many similarities in the estimation methods of the two model parameters. Therefore, the research conclusions on parameter estimation methods for ordinary logistic models are also applicable to parameter estimation methods for quadratic logistic models. The differences are that the process of parameter estimation in the quadratic logistic model is more complex, stability of estimated results is relatively poor, and the ways are relatively few. Therefore, this method has stricter requirements for data quality and sample path strength.

## Acknowledgement:

This research was sponsored by the National Natural Science Foundation of China (Grant No. 42171192). The support is gratefully acknowledged.

## References

Batty M (1991). Generating urban forms from diffusive growth. *Environment and Planning A*, 23: 511-544

Batty M, Longley PA (1994). *Fractal Cities: A Geometry of Form and Function*. London: Academic Press

Benguigui L, Czamanski D, Marinov M, Portugali J (2000). When and where is a city fractal? *Environment and Planning B: Planning and Design*, 27(4): 507–519

Bertalanffy L von (1968). *General System Theory: Foundations, Development, and Applications*. New York: George Braziller

Cadwallader MT (1996). *Urban Geography: An Analytical Approach*. Upper Saddle River, NJ: Prentice Hall

Chen YG (2009). Carrying capacity estimation of logistic model in population and resources prediction by nonlinear autoregression. *Journal of Natural Resources*, 24(6): 1105-1113 [In Chinese]

Chen YG (2012). Fractal dimension evolution and spatial replacement dynamics of urban growth. *Chaos, Solitons & Fractals*, 45 (2): 115-124

Chen YG (2018). Logistic models of fractal dimension growth of urban morphology. *Fractals*, 26(3): 1850033

Chen YG, Huang LS (2019). Modeling growth curve of fractal dimension of urban form of Beijing. *Physica A: Statistical Mechanics and its Applications*, 523: 1038-1056

Feng J, Chen YG (2010). Spatiotemporal evolution of urban form and land use structure in Hangzhou, China: evidence from fractals. *Environment and Planning B: Planning and Design*, 2010, 37(5): 838-856

Fisher JC, Pry RH (1971). A simple substitution model for technological change. *Technological Forecasting and Social Change*, 3: 75-88

Frankhauser P (1994). *La Fractalité des Structures Urbaines (The Fractal Aspects of Urban Structures)*. Paris: Economica

Frankhauser P (1994). *La Fractalité des Structures Urbaines (The Fractal Aspects of Urban Structures)*. Paris: Economica

Frankhauser P (1998). Fractal geometry of urban patterns and their morphogenesis. *Discrete Dynamics in Nature and Society*, 2(2):127-145

Frankhauser P (1998). The fractal approach: A new tool for the spatial analysis of urban agglomerations. *Population: An English Selection*, 10(1): 205-240

Hermann R, Montroll EW (1972). A manner of characterizing the development of countries. *PNAS*, 69(10): 3019-3024

Jiang SG, Liu DS (2012). Box-counting dimension of fractal urban form: stability issues and measurement design. *International Journal of Artificial Life Research*, 3 (3), 41-63

Jiang SG, Zhou YX (2006). The fractal urban form of Beijing and its practical significance. *Geographical Research*, 25(2): 204-213[In Chinese]

Karmeshu (1988). Demographic models of urbanization. *Environment and Planning B: Planning and Design*, 15(1): 47-54

Longley PA, Batty M, Shepherd J. The size, shape and dimension of urban settlements. *Transactions of the Institute of British Geographers (New Series)*, 1991, 16(1): 75-94

Man XM, Chen YG (2020). Fractal-based modeling and spatial analysis of urban form and growth: a case study of Shenzhen in China. *ISPRS International Journal of Geo-Information*, 9(11): 672

Morrill R, Gaile GL, Thrall GI (1988). *Spatial Diffusion*. Newbury Park, CA: SAGE publications

Rao DN, Karmeshu, Jain VP (1989). Dynamics of urbanization: the empirical validation of the replacement hypothesis. *Environment and Planning B: Planning and Design*, 16(3): 289-295

Sambrook RC, Voss RF (2001). Fractal analysis of US settlement patterns. *Fractals*, 9(3): 241-250

Shen G (2002). Fractal dimension and fractal growth of urbanized areas. *International Journal of Geographical Information Science*, 16(5): 419-437

Sun J, Southworth J (2013). Remote sensing-based fractal analysis and scale dependence associated with forest fragmentation in an Amazon tri-national frontier. *Remote Sensing*, 5(2), 454-472

United Nations (2004). *World Urbanization Prospects: The 2003 Revision*. New York: U.N. Department of Economic and Social Affairs, Population Division

White R, Engelen G (1993). Cellular automata and fractal urban form: a cellular modeling approach to the evolution of urban land-use patterns. *Environment and Planning A*, 25(8): 1175-1199

Zhou YX (1995). *Urban Geography*. Beijing: The Commercial Press [In Chinese]